\documentclass[review]{elsarticle}

\usepackage{hyperref}
\usepackage{amsmath,amssymb} 
\usepackage{graphicx}
\usepackage{wrapfig}
\usepackage[version=4]{mhchem}
\usepackage{xcolor}

\usepackage[normalem]{ulem}
\newcommand{\m}{\ensuremath{_{matrix}}}
\newcommand{\p}{\ensuremath{_{prec}}}
\newcommand{\ten}[1]{\boldsymbol{#1}}

\journal{Journal of \LaTeX\ Templates}

\begin{document}

\begin{frontmatter}

\title{ Ab initio Modelling of the
Early Stages \\of Precipitation in Al-6000 Alloys} %

\author[1]{Daniele Giofr\'e\corref{corresponding}}%
\cortext[corresponding]{Corresponding author: daniele.giofre@epfl.ch}

\author[2]{Till Junge}

\author[2]{W. A. Curtin}

\author[1]{Michele Ceriotti}

\address[1]{Laboratory of Computational Science and Modeling, Institute of Materials, 
\'Ecole Polytechnique F\'ed\'erale de Lausanne, 1015 Lausanne, Switzerland}

\address[2]{Laboratory for Multiscale Mechanics Modeling, Institute of Mechanical Engineering, EPFL, 1015 Lausanne, Switzerland}

\begin{abstract}
Age hardening induced by the formation of 
(semi)-coherent precipitate phases
is crucial for the processing and final 
properties of the widely used Al-6000 alloys. 
Early stages of precipitation are particularly 
important from the fundamental and technological 
side, but are still far from being 
fully understood. Here, 
an analysis of the energetics of nanometric
precipitates of the meta-stable 
$\beta''$ phases is performed, identifying the
bulk, elastic strain and interface energies that
contribute to the stability of a
nucleating cluster.
Results show that needle-shape precipitates 
are unstable to growth even at the smallest size $\beta''$ formula
unit, i.e. there is
no energy barrier to growth.
The small differences between different compositions points toward
the need for the study of possible precipitate/matrix interface
reconstruction.
A classical semi-quantitative nucleation theory approach including elastic strain energy 
captures the trends in precipitate energy 
versus size and composition.  This
validates the use of mesoscale models
to assess stability and interactions of precipitates.
Studies of smaller 3d clusters also show stability relative to the solid solution state,
indicating that the early stages of precipitation may be diffusion-limited.  Overall,
these results demonstrate the important interplay among composition-dependent bulk, interface,
and elastic strain energies in determining nanoscale precipitate stability and growth.
\end{abstract}


\end{frontmatter}


%
\section{Introduction}

Pure aluminum is  lightweight metal that has
little strength or resistance to plastic deformation. 
Alloying aluminum introduces either solutes or the formation of nanometric precipitates 
that hinder the motion of dislocations, 
thereby dramatically improving the mechanical properties
\cite{murayama_pre-precipitate_1999, edwards_precipitation_1998, ringer_microstructural_2000}. 
A major alloy class used in the automotive industry is the Al-6000 series that contains
silicon and magnesium in the range of 
0.4--1 wt\% with a Si/Mg ratio larger than one.  
In the initial stages of processing at elevated temperatures, 
the alloy is a supersaturated solid solution (SSSS), with the solutes randomly dispersed in the Al matrix.
After quenching to lower temperatures, the solutes aggregate to form nanometer-sized precipitates 
(e.g  Guinier-Preston (GP) zones, metastable phases, or stable phases, depending on the 
thermal history).  The time evolution of precipitate nucleation and growth is accompanied by a concomitant
mechanical strengthening, referred to as age-hardening. 
Furthermore, precipitation proceeds through a sequence of competing phases 
that differ in composition, morphology, thermodynamic stability, and kinetics of growth and dissolution, 
as well as in the contributions to the mechanical properties~\cite{ravi-wolv04am,marioara_influence_2005}.
Control of the kinetics of age-hardening is crucial for the optimization of 
the final mechanical properties.

In commercial 6000-series Al alloys, precipitation commences at room temperature 
shortly after quenching, and this ``natural aging'' is undesirable.  Subsequent
``artificial aging'' at elevated temperature is then used to achieve
the desired precipitate type(s) and sizes.  The most effective hardening 
conditions are obtained in the early stages of precipitation, where 
fully-coherent GP zones coexist with the semi-coherent $\beta''$ 
phase~\cite{takeda_stability_1998}, which forms needle-shaped
precipitates 200-1000 {\AA} in length and $\approx$ 60 {\AA} in 
diameter \cite{andersen_crystal_1998, zandbergen_data_2015}.
High-resolution electron microscopy and 
quantitative electron diffraction \cite{marioara_influence_2003, andersen_crystal_1998} studies
have revealed that the $\beta''$ phase is characterized by a Mg/Si ratio close to 1 
but with different possible stoichiometries that include \ce{Mg5Si6}, \ce{Mg4Al3Si4}, \ce{Mg5Al2Si4}.
Recent first-principles calculations have predicted that the 
latter composition is the most stable\cite{niniveIpaperthesis2014}.
While considerable progress has been made in understanding
the structure of the $\beta''$ phase, and the behavior of the 
SSSS~\cite{poga+14prl}, little is known on the early stages of 
the aging mechanism, and in particular on the thermodynamics of 
the initial clustering of solutes to form 
the precipitate~\cite{murayama_pre-precipitate_1999, edwards_precipitation_1998, marioara_atomic_2001, marioara_influence_2003}.
Such knowledge is crucial
to gain better control over the balance between natural and 
artificial aging. 

In the present work we study the energetics of nanoscale precipitates 
using \emph{ab initio} electronic structure methods so as to identify the different
contributions to the thermodynamic in-situ precipitation energetics.  
We compute the energy contributions due to the precipitate formation energy,
the precipitate/matrix interface energies, and the elastic energy due to lattice and elastic
mismatch between precipitate and matrix.  We show that these contributions semi-quantitatively
capture the total energy of in-situ precipitates as a function of precipitate size.
Our results demonstrate that -- down to the size of a single formula unit of the $\beta''$ phase,
fully encapsulated in the Al matrix -- the precipitate growth process can proceed without 
energetic barriers.  Since the nucleation process of the $\beta''$ phase has nearly zero barrier,
control of precipitation kinetics should focus 
on aggregates of atoms of even smaller size.

The remainder of this paper is organized as follows.  In Section~\ref{sec:details} we 
describe the details of our \emph{ab initio}
simulations.  In Section~\ref{sec:bulk} we 
report a few benchmarks on the bulk properties
of the different stoichiometries proposed for the
$\beta''$ phases. In Section~\ref{sec:insitu} we
discuss a classical-nucleation-theory (CNT) model of precipitate
stability, including surface energies and the 
continuum elasticity model of lattice mismatch
relaxation, and compare with DFT results 
for needle-like precipitates.  In 
Section~\ref{sec:threed} we present \emph{ab 
initio} simulations of fully-encapsulated clusters. We finally draw conclusions.

\section{Computational details}
\label{sec:details}
Density functional theory (DFT) has been shown to provide reliable 
energetics for aluminum and its 
alloys~\cite{niniveIpaperthesis2014, derlet_first-principles_2002, hasting_composition_2009,poga+14prl}.
We have used self-consistent DFT as implemented in 
the Quantum ESPRESSO (QE) package\cite{gian+09jpcm}.
We used a gradient corrected exchange and correlation 
energy functional (PBE)\cite{perd+96prl}, together
with a plane-waves expansion of Kohn-Sham orbitals and
electronic density, using ultra-soft pseudopotentials
for all the elements involved~\cite{vand90prb, kresse_ultrasoft_1999, materialcloud}.
All calculations were performed with a $k$-point
sampling of the Brillouin zone using a grid 
density of $\approx 5\cdot10^{-6}$ \AA$^{-3}$ and a 
Mokhorst-Pack mesh\cite{monk-pack76prb}. 
The plane-wave cut-off  energy 
was chosen to be 35 (280) Ry
for the wavefunction (the charge density) when evaluating the 
energetics of defects (i.e. for computing 
formation, surface, and precipitation energies).
Test calculations performed at larger cutoffs 
showed that these parameters are sufficient to converge
the atomization energy of Al at a level of 0.3 meV/atom. 
Cutoffs were increased to 50 (400) Ry so as to 
converge the value of the elastic constants
to an error below 1 GPa. Comparison with previous literature results, where available, will
be presented below.

\section{Bulk properties of matrix and precipitate phases}
\label{sec:bulk}
Bulk properties (lattice structure, lattice constants, elastic constants) of Al and the
various $\beta''$-precipitates studied here have been previously computed in the literature.
Here, we present our results as a means of benchmarking our methods, verifying
literature results, and most importantly obtaining reference values that are
fully consistent with our computational details -- which is crucial to evaluate
the energy differences that determine surface and defect energies. 

For bulk \emph{fcc} Al, we computed the lattice parameter to be 
4.057 \AA{}, in excellent agreement with the experimental value and with previous 
modelling using the same functional~\cite{davey_precision_1925, tambe_bulk_2008}.
These lattice parameters are used throughout our study to build supercells 
representing the Al matrix.
All of the $\beta''$ phases we consider can be described by a monoclinic cell containing two 
formula units (f.u.).  We consider three compositions,  \ce{Mg5Si6}, \ce{Mg5Al2Si4} and 
\ce{Mg4Al3Si4}, as shown in Figure \ref{tab:lattice}.   We computed the
crystal structures of these $\beta''$-precipitates starting from the geometries 
proposed in previous works~\cite{andersen_crystal_1998}.
The equilibrium lattice parameters and monoclinic angles
are shown in Table~\ref{tab:lattice}, and agree well with
existing literature~\cite{ravi-wolv04am}.
Inside the Al matrix, the main crystallographic directions (lattice vectors) of the
precipitate are aligned with those in the \emph{fcc} lattice of aluminum as follows:
\begin{equation}
[100]_{\beta''} \parallel [203]_{\ce{Al}} \qquad [010]_{\beta''} \parallel [010]_{\ce{Al}}  \qquad [001]_{\beta''} \parallel [\bar{3}01]_{\ce{Al}}.
\label{eq:interfacerel}
\end{equation}
The ideal monoclinic unit cell can be deformed, relative to the 
fully relaxed structures, to substitute for 22 Al atoms.  The corresponding
lattice vectors and lattice constants of the 22-atom Al are shown Table~\ref{tab:lattice}.
The difference between the ideal monoclinic unit cell and the 22-atom Al unit cell uniquely
determines the misfit strain tensor of the precipitate in the Al lattice, which will be used below
to determine the corresponding elastic energy of precipitates in the matrix.

\begin{table}[tpbh]
  \includegraphics[width=0.25\columnwidth]{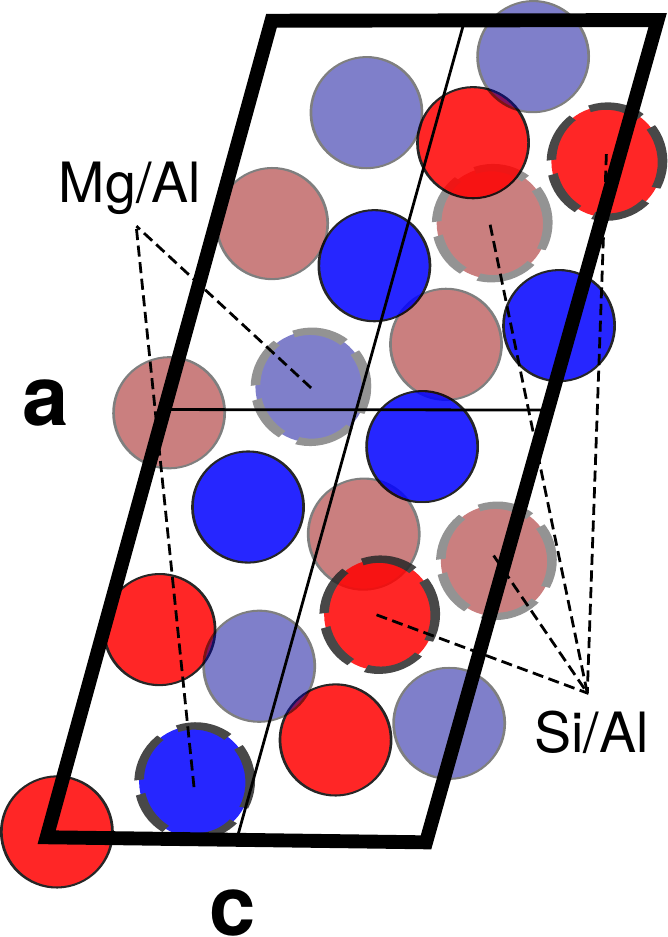}
\qquad  
\begin{tabular}[b]{   c | c c c c  }     
\hline\hline  Composition   & a [\AA]  & b  [\AA] & c [\AA]  & $\beta$ [$^\circ$]\\  
\hline  \ce{Mg5Si6}   & 15.14   & 4.08  & 6.93  & 109.9\\  
  \ce{Mg5Al2Si4}   & 15.33  & 4.05 & 6.84  & 106.0\\  
  \ce{Mg4Al3Si4}   & 15.13  & 4.12 & 6.65  & 106.6\\  
  Matrix\cite{andersen_crystal_1998}  & 14.63  & 4.06 & 6.41 & 105.3\\  
\hline\hline  \end{tabular}
\caption{\label{tab:lattice}(Left) A view along the $\mathbf{b}$ lattice vector of the monoclinic 
unit cell of $\beta''$ phases. The red and blue
circles represent Si and  Mg atoms, respectively, while the different shading 
indicates the position of the atoms at a height of 
zero and $|$\textbf{b}$|$/2 along the \textbf{b} 
vector. 
Circles with dashed outline indicate the 
atoms that can be substituted to obtain
the three $\beta''$ stoichiometries 
(that is, \ce{Mg5Si6}, \ce{Mg5Al2Si4} and 
\ce{Mg4Al3Si4})\cite{niniveIpaperthesis2014}. (Right) The fully-relaxed bulk lattice 
parameters of the $\beta''$ phases, compared 
with those that correspond to an ideal 
embedding within the Al matrix.}
\end{table}

We computed the elastic constants of all bulk phases by evaluating the stresses generated by
small displacements of the unit cell around the equilibrium structure.  A suitable set of displacements
was used, and the stresses were then modelled as a linear function of the displacements
to obtain the elastic constants ~\cite{nielsen_first-principles_1983}.  The elastic constants 
for bulk Al and for the three $\beta''$ phases studied here are shown in Table~\ref{tab:elastic_constants},
and were computed according to a reference system consistent with the Al matrix, as shown in Fig.~\ref{fig:orientation}.
Our values are in good agreement with available experimental
values~\cite{elasticonstant2002} and previous computations
~\cite{bercegeay_first-principles_2005, niniveIpaperthesis2014, yu_calculations_2010}.

\begin{table}
\centering{\begin{tabular}{  c  c  c  c   c  c  c  c   }     
\hline \hline   \multicolumn{2}{c|}{[GPa]} & $C_{11}$ & $C_{22}$ & $C_{33}$ & $C_{44}$ & $C_{55}$ & $C_{66}$ \\
\hline    \multicolumn{2}{c|}{Al} &  \multicolumn{3}{c}{106.1 (114.3)} & \multicolumn{3}{c}{ 31.9 (31.6) } \\ 
  \multicolumn{2}{c|}{\ce{Mg_5Si_6}} & 98.4 &  84.6  & 88.0 & 21.9 & 29.1 & 51.2 \\
 \multicolumn{2}{c|}{\ce{Mg_5Al_2Si_4}} & 107.1 & 94.7  & 99.1 & 26.9 & 36.3 & 49.4  \\ 
 \multicolumn{2}{c|}{\ce{Mg_4Al_3Si_4}} & 106.7 & 96.5  & 97.1 & 25.9 & 35.6 & 46.3 \\          
\hline \hline \hphantom{$C_{12}$}  &   $C_{12}$ & $C_{13}$ & $C_{23}$ & $C_{15}$ & $C_{25}$ & $C_{35}$ & $C_{46}$  \\   
\hline   &  \multicolumn{3}{c}{ 55.9 (61.9)}& \multicolumn{4}{c}{0.}     \\  
 & 50.0 & 47.7 & 45.7 & 8.2 & 5.8 & 5.4 & -10.1    \\
 & 40.3 & 45.6 & 43.0 & -13.1 & 4.3 & 11.9   & 5.4   \\ 
 & 46.5 & 48.0 & 48.8 & 9.3 & 5.7 & 9.3 & 6.3    \\   
 \hline\hline   \end{tabular} }
\caption{Elastic constants obtained by a 
linear fit of ab-initio stress tensors for 
small cell deformations. 
The values in parentheses are the 
experimental ones, extrapolated to 0K.\cite{elasticonstant2002} 
 \label{tab:elastic_constants}}
\end{table}

In order to define a reference state for the thermodynamics of the precipitates
we define the solid solution energies as
\begin{align}
E^{ss}_{\ce{Al}} &= E^{tot}_{\ce{Al}_{M}}/M\\
E^{ss}_{\ce{x}} &= E^{tot}_{\ce{Al}_{M-1}\ce(x)}-(M-1) E^{ss}_{\ce{Al}},
\end{align}
for $x = \ce{Si,Mg}$. Here, $E^{tot}_{\ce{Al}_{M}}$ and $E^{tot}_{\ce{Al}_{M-1}\ce(x)}$ 
are the total energies of a bulk-Al supercell containing $M$ Al atoms and $(M-1)$ Al atoms
and $1$ atom of $x = \ce{Si,Mg}$, respectively.  The energy $E^{tot}_{\ce{Al}_{M-1}\ce(x)}$ is computed
using a single solute in a 4x4x4 unit periodic cell with the cell volume held fixed.  
The cell develops a 
small pressure due to the misfit volume of the solute, but this contribution to the energy 
is negligible for the large cell size used.

The formation energy for a precipitate can then be defined as the total energy of a precipitate formula
unit relative to that of the total energies of the precipitate atoms in the solid solution state.
Thus, the formation energy is
\begin{equation}
E_{form}=\frac{1}{2}\, E^{tot}_{\beta''}
- \sum_{x=\ce{Al, Si, Mg}} 
n_{x}\cdot E^{ss}_{x},
\label{eq:formation}
\end{equation}
where $E^{tot}_{\beta''}$ is the (DFT) total energy of 
a fully-relaxed unit cell of the $\beta''$ phase containing 
22 atoms (2 formula units), $n_{x}$ is the number of atoms of element $x$ in one
formula unit, and $E^{ss}_{x}$ is the energy of solute $x$ in the (dilute)
solid solution state. 
Knowing all the terms in eq. \ref{eq:formation}, we can compute 
the formation energies of the three proposed $\beta''$-phase compositions
as shown in Table \ref{tab:valori}).   The precipitates are strongly favorable,  
with negative formation energies in excess 
of -2eV/f.u., or greater than -0.2 eV/atom on average.  Precipitate formation
is thus thermodynamically highly preferable relative to the solid solution state.

\begin{table}
  \centering
  {\small
  \begin{tabular}[b]{   c | c | c c c  | c c c c  }    
    \hline\hline
             & $E_{form}$  & \multicolumn{3}{|c|}{$\gamma$ [meV/\AA$^2$]}  & \multicolumn{4}{|c}{$E_{strain}$ [meV/f.u.] (size: f.u./\emph{l.u.})}\\ 
                      &  [eV/f.u.]  & \multicolumn{3}{|c|}{\emph{[mJ/m$^2$]}  }         & dilute & $N: 1$ & $4$ & $16$  \\
                      &             &    &    &           & $1\times1$ & $1\times1$ & $2\times2$ & $4\times4$   \\
                      &              & A   & B   & C      &  \emph{96$\times$96} & \emph{5$\times$5} & \emph{7$\times$7} & \emph{12$\times$12} \\\hline
    \ce{Mg_5Si_6}     & -2.607      & 8.36 & 21.1 & 2.69      & 140  & 171 & 203 & 223  \\ 
                    & &  \emph{134} &  \emph{338} &  \emph{43.1} & & & & \\
    \ce{Mg_5Al_2Si_4} & -2.769      & 11.8 & 23.5 & 9.11     & 128  & 161 & 198 & 223  \\
    & &  \emph{189} &  \emph{376} &  \emph{146} & & & & \\
    \ce{Mg_4Al_3Si_4} & -2.366     & 10.1 & 20.4 & 8.24      & 74 %
                                                                 & 89%
                                                                           & 106 & 117  \\     
                    & &  \emph{162} &  \emph{327} &  \emph{132} & & & & \\
    \hline\hline  \end{tabular}
  }
\caption{Bulk, strain, and surface-energy terms
computed for the three stoichiometries of the
$\beta''$ phase which we considered in this
study.
The elastic strain energy $E_{strain}$ has been computed for the dilute case and for the three periodic cases with varying numbers of formula units ($N$). The precipitate size is reported in formula units (f.u.) and the matrix size in fcc lattice unit cells (l.u.).
}
\label{tab:valori}
\end{table}

\section{In-situ precipitates} 
\label{sec:insitu}

Bulk properties provide important information on the thermodynamic driving forces for
precipitation, but are incomplete for understanding in-situ precipitation nucleation and growth.
The system of precipitate plus matrix has additional energetic contributions from
the precipitate/matrix interfaces, precipitate/matrix lattice and elastic constant mismatches that give
rise to elastic energies when the precipitate is coherent, and precipitate/matrix edge and corner
energies.  All of these additional contributions determine the total thermodynamic driving 
force for precipitate growth as a function of precipitate size, shape, and density.
While not addressed here, the elastic interactions between precipitates at finite densities 
also influences their spatial arrangement and orientation~\cite{li_computer_1998, luo_stress/strain_2014, fu_effects_2014}.

We thus need to predict the size, shape, and energy of a critical precipitate
nucleus.  At some critical precipitate size, the precipitate becomes thermodynamically unstable 
to further growth, i.e. increasing size leads to decreasing total energy.  Below the critical precipitate size,
the precipitate is unstable and should re-dissolve in the solid solution.  Here, we take
a model based on classical nucleation theory (CNT) to assess the precipitate stability as a function of size, shape and 
density (which influences the elastic energy).  In this analysis, we ignore edge and corner
energies.  Also assuming, for the moment, a low density of precipitates, the total energy
of a precipitate containing N formula units, relative to the SSSS, can be written as
\begin{equation}
E_{prec}(N) = N E_{form} + N E_{strain} + E_{surf}(N).
\label{eq:tdmodel}
\end{equation}
There are two new terms in Eq.~\ref{eq:tdmodel}.  First, there is the elastic strain 
energy $E_{strain}$ due to  the lattice and elastic mismatch between 
the precipitate and the Al-matrix per $\beta''$ formula unit for a single precipitate in an infinite
matrix (the dilute limit).  Second, there is the surface (interface) energy $E_{surf}$ of the 
precipitate, which will depend on both the size and the shape of the nucleus. 
In order to evaluate the precipitation energy, we first obtain quantitative values for the strain and 
interface energies. Then, we will make predictions for the thermodynamics in the dilute limit.
Finally, we will perform DFT studies of in-situ precipitates and compare the DFT energies
versus the CNT model, adapted to the geometry of the DFT supercells.

\subsection{\texorpdfstring{Interface energies for $\beta''$ precipitates}{Interface energies for beta'' precipitates}}

\begin{figure}[bthp]
\centering  \includegraphics[width=0.95\columnwidth]{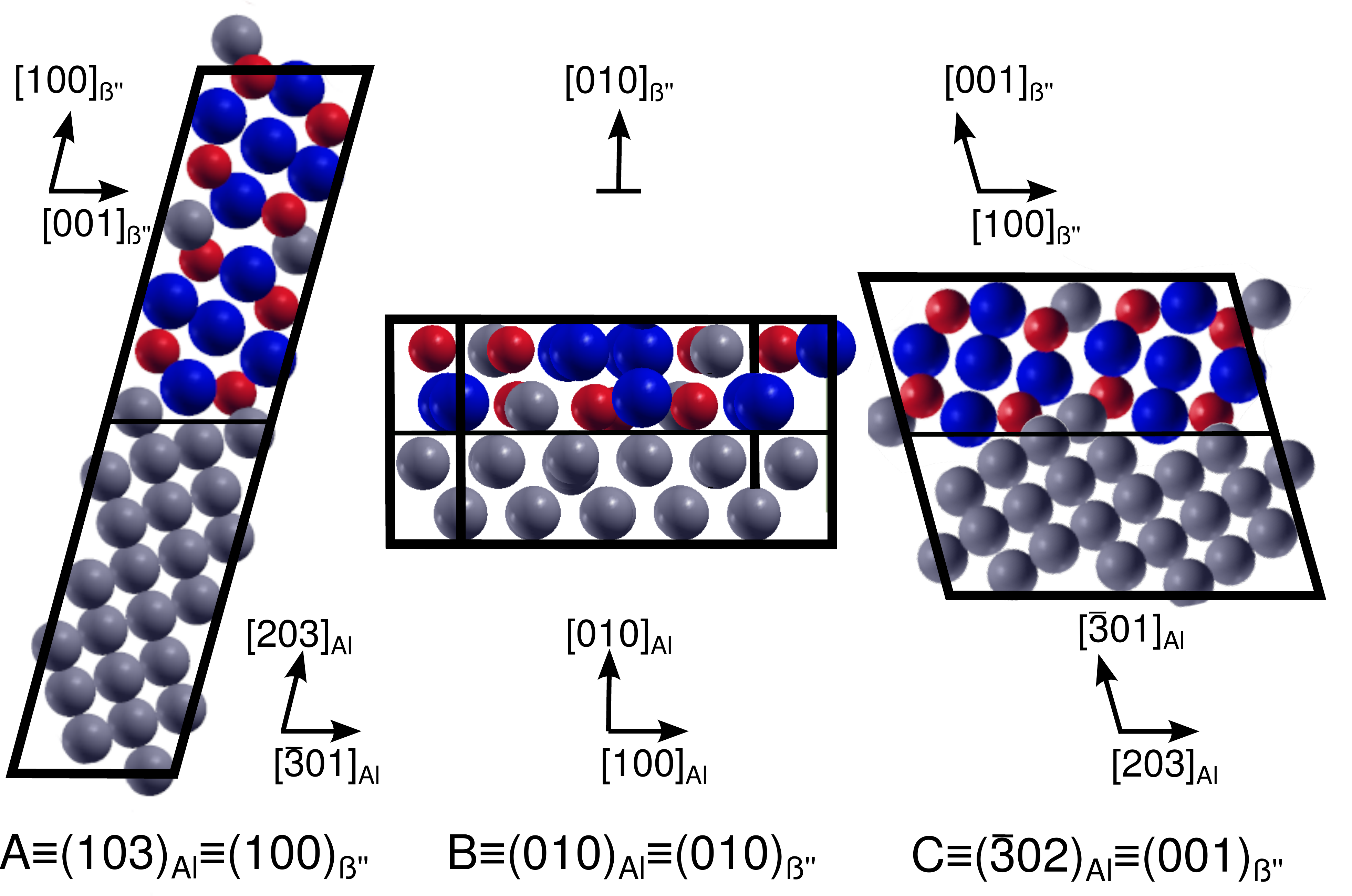}
\caption{\label{fig:surfaces} 
We considered three orientations for the
interfaces the between $\beta''$ precipitate 
and the Al matrix: the A 
 orientation (left), the B orientation (center), and  the C orientation 
 (right).  
While we chose to represent only the 
\ce{Mg5Al2Si4} phase, for simplicity, 
the other two compositions can be obtained 
by performing the substitutions indicated 
in Tab. \ref{tab:lattice}.}
\end{figure}

Based on TEM analyses \cite{yao_tem_2001, marioara_influence_2003, marioara_influence_2005}, 
and the correspondence between $\beta''$ structure and the closely-related 22-atom Al unit 
that accommodates one precipitate unit cell, we study three interface orientations
as shown in Figure~\ref{fig:surfaces}.  The orientations are denoted
$\text{A}\equiv (103)_{\ce{Al}}\equiv (100)_{\beta''}$, 
$\text{B}\equiv (010)_{\ce{Al}}\equiv (010)_{\beta''}$,
and
$\text{C}\equiv (\bar{3}02)_{\ce{Al}}\equiv (001)_{\beta''}$.
Given the relatively complex structure of the
$\beta''$ phase, there are many possible ways to terminate the 
precipitate.  Previous computational studies of the $\beta''$-\ce{Mg5Si6}/$\alpha$-Al interface have found that
the associated surface energies can change significantly 
between different choices \cite{wang_first-principles_2007}. 
To compare with previous studies of finite-size precipitates, 
we chose the interfaces used in Ref.~\citenum{niniveIpaperthesis2014}.
Figure~\ref{fig:surfaces} shows only one monoclinic unit 
cell of the precipitate and one for the matrix for \ce{Mg5Al2Si4} but all three
compositions were studied, and simulations were performed with much larger
supercells of sizes $4 \beta'' + 4 \ce{Al} $ unit cells for the A 
 orientation,  $6 \beta'' + 6 \ce{Al} $ unit cells for 
 the B orientation, and  $6 \beta'' + 6  \ce{Al} $ unit cells for the C orientation. 

Since the precipitate and matrix have a structural mismatch, the total energy
computed in a given simulation cell includes an elastic deformation energy.  This 
energy must be computed independently and subtracted from the total energy
obtained in the interface simulation to estimate the specific interface energy $\gamma_{\Lambda= A,B,C}$.  
First, we compute the energy per formula unit of the partially-relaxed  $\beta''$ phase.
For each interface orientation, we define 
$E_{\Lambda}^{\beta''}$ as the energy per formula
unit of a $\beta''$ cell that is fully coherent
with the Al matrix in the $\Lambda$ plane, and 
relaxed in the orthogonal direction. 
We then prepared an interface between the Al matrix and the $\beta''$ phase, once again
fixing the dimensions parallel to the interface
to be fully coherent with the matrix, and relaxing it in the orthogonal direction. 
The interface energy can then be obtained 
from the total energy of this supercell $E^{sc}_\Lambda$ as 
\begin{equation}
\gamma_{\Lambda= A,B,C} =\frac{\left( E^{sc}_{\Lambda}-n_{Al} E^{SS}_{\ce{Al}} - n_{\beta''} E_{\Lambda}^{\beta''}\right)}{2 S^\Lambda_{supercell}},
\end{equation}
where $S^\Lambda_{supercell}$ is the cross-section
of the simulation supercell corresponding to the 
orientation of the interface, $n_{Al}$ is the number of Al atoms in the matrix, 
and $n_{\beta''}$ is the number of $\beta''$ formula units inside the supercell.
The computed surface energies for each orientation are shown in Table~\ref{tab:valori}.

As previously noted~\cite{wang_first-principles_2007},
the $B$ surface energy is relatively large but the anisotropy is not sufficient 
to fully explain the observed needle-shaped habit of the precipitates. 
Given the large range of values observed for
different terminations~\cite{wang_first-principles_2007}, 
a change in composition or some degree of interface reconstruction
may significantly lower the energies of the A and C interfaces, leading to 
larger anisotropy.  For instance, we obtain a considerably lower surface energy for the C interface in \ce{Mg5Si6} than any of the values reported in
Ref.~\citenum{wang_first-principles_2007}. For this specific case -- that is associated with a relatively large mismatch in the unit cells between the $\beta''$ phase and the matrix -- we observe significant relaxation of atoms at the interface, extending for several layers in the bulk, that was probably not captured fully in the smaller supercells\footnote{Calculations in Ref.~~\citenum{wang_first-principles_2007} used 44+44 atoms supercells, while our calculations for the C interface contained 132+132 atoms. We verified that when using a supercell with 66+66 atoms the surface energy for \ce{Mg5Si6}(C) increased to 63 mJ/m$^2$, getting closer to previous results.} used in Ref.~\citenum{wang_first-principles_2007}. 
The issue of interface energies of $\beta''$ phases in Al
thus merits further study. 

\subsection{\texorpdfstring{Elastic strain energies of needle-like $\beta''$ precipitates}{Elastic strain energies of needle-like beta'' precipitates}}

\begin{figure}[htb]
  \centering
  \includegraphics[width=0.5\textwidth]{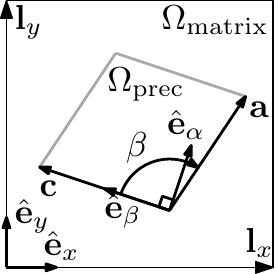}
  \caption{Schematic of the computational domain and definition of frames of reference. The directions of the vectors $\vec c$ and $\vec a$ are drawn as defined by \eqref{eq:interfacerel}, vectors $\vec b$ and $\hat{\vec e}_z$ point out of the paper (not depicted). The global frame of reference is $\hat{\vec e}_x$-$\hat{\vec e}_y$-$\hat{\vec e}_z$ while the elastic constants listed in  Table~\ref{tab:elastic_constants} are measured in the material frame $\hat{\vec e}_\alpha$-$\hat{\vec e}_\beta$-$\hat{\vec e}_z$.}
  \label{fig:orientation}
\end{figure}

During the aging process,  $\beta''$ precipitates show a 
strongly anisotropic habit, extending along
the $b\equiv[010]$ direction forming needle-like semi-coherent
particles.  The lattice mismatch between Al and $\beta''$ along the crystallographic
$b$ direction is also quite small.  For this reason, two-dimensional slices 
along the $a,c$ axes of the precipitate capture the
main contributions to the energetics of large precipitates,
and have already been studied to characterize both the 
energetics and elastic deformation of the matrix in this
regime~\cite{niniveIIpaperthesis2014, wang_first-principles_2007}.
To compute the 
magnitude of the elastic strain energy contribution for such a two-dimensional slice, we will use anisotropic continuum elasticity. 
The boundary value problem is formulated to correspond to the direct DFT studies below.
We study a periodic two-dimensional plane-strain problem with a fully three-dimensional eigenstrain within the precipitate due to the misfit between the precipitate and the matrix.  Figure~\ref{fig:orientation} shows a schematic of the geometry with the relevant
coordinate axes.

The Al matrix $\Omega\m$ is modeled as linearly elastic, 
\begin{equation}
  \label{eq:constitutive_aniso}
  \boldsymbol\sigma = \ten C\m\boldsymbol\epsilon\quad\text{ in }\Omega\m,
\end{equation}
where $\boldsymbol\sigma$ and $\boldsymbol\epsilon$ are the Cauchy stress and strain tensors and $\ten C\m$ is the anisotropic fourth-order stiffness tensor of the matrix expressed in the global frame of reference $\hat{\vec e}_x$-$\hat{\vec e}_y$-$\hat{\vec e}_z$
aligned with the cubic lattice vectors of the pure aluminum matrix. The precipitate $\Omega\p$ is also linearly elastic, but
with an additional eigenstrain $\bar{\boldsymbol\epsilon}$ relative to the reference Al lattice
that accounts for the size and shape misfit of the precipitate,
\begin{equation}
  \label{eq:constitutive_aniso_eigen}
  \boldsymbol\sigma = \ten C\p(\boldsymbol\epsilon-\bar{\boldsymbol\epsilon})\quad\text{ in }\Omega\p.
\end{equation}
Determination of the eigenstrain $\bar{\boldsymbol\epsilon}$ and the rotation of the stiffness tensor $\ten C\p$ into the global frame of reference are described in the  \ref{sec:eigenstrain}.  

As a plane-strain problem, there is zero out-of-plane displacement $u_z = 0$.   Therefore the total strain tensor has $\epsilon_{xz}=\epsilon_{yz}=\epsilon_{zz} = 0$.  The eigenstrain $\bar{\boldsymbol\epsilon}$ retains these components, however, so
that the effects of the mismatch in the $z$ direction are included.  We impose periodic Dirichlet boundary conditions on the 
displacement $\vec u$ in the horizontal and vertical directions
\begin{equation}
  \label{eq:bndcond}
  \vec u(\vec x) = \vec u(\vec x + n\vec l_x + m\vec l_y), \quad n,m \in \mathbb{Z},\quad \forall \vec x \in \partial\Omega,
\end{equation}
where $\vec l_x$ and $\vec l_y$ are the vectors linking the bottom left corner to the bottom right and the top left, respectively.
We fix an arbitrary point $u(\vec x_p) =0$ to exclude solid body motion.
The static equilibrium stress and strain fields throughout the body are then determined by solving the standard equilibrium equation
$\boldsymbol\nabla\cdot\boldsymbol\sigma = \vec0$.
With the computed stess field, the strain fields are obtained from the constitutive models above and 
the elastic strain energy (per unit length in the out-of-plane direction) $E_\text{strain}$ is then computed as
\begin{equation}
  \label{eq:strain_energy}
  E_{strain} = \frac{1}{2}\left(\int_{\Omega_{matrix}}{\boldsymbol\epsilon\ten C_{matrix}\boldsymbol\epsilon \,\mathrm d\Omega} + \int_{\Omega_{prec}}{\ten (\boldsymbol\epsilon - \bar{\boldsymbol\epsilon})C_{prec}(\boldsymbol\epsilon - \bar{\boldsymbol\epsilon})\,\mathrm d\Omega}\right)
\end{equation}
Note that the energy per unit length is independent of absolute model size and so the energy only depends on the size of the precipitate
relative to the size of the computational cell, or equivalently on the area fraction (equal to the volume fraction) of the precipitate.

The boundary value problem 
is solved using the finite-element method (see \ref{sec:elastic_calculations}).
Note that, although the problem is nominally two-dimensional (plane-strain), 
the evaluation of the elastic strain energy remains fully three-dimensional due to the eigenstrain $\bar{\boldsymbol\epsilon}$.  
Using the above implementation, we first computed the elastic strain energy per formula unit in the dilute limit where interactions among
precipiates are negligible.  This is done by using one formula
unit in a cell of 96 x 96 fcc unit cells, and the results are shown as the ``dilute" limit in Table~\ref{tab:valori}.  The elastic energies
are small compared to the chemical energies, but are not small compared to differences in energies among precipitate compositions.

\subsection{\texorpdfstring{In-situ energetics of dilute needle-like $\beta"$ precipitates}{In-situ energetics of dilute needle-like beta" precipitates}}

Having evaluated separately the bulk, 
surface, and elastic relaxation energies for a
needle-like precipitate of the $\beta''$ phases,
we can then proceed to estimate the overall
energetics of a nucleus. Assuming for simplicity
the surface area of the interfaces to be that of the 
matrix-coherent unit cell
(that is 26.02 \AA$^2$ for each formula unit 
along the A facets, and 29.7  \AA$^2$ for each
formula unit along the C facets)  we find that a needle-like precipitate
with a cross-section of a single formula has
already a negative formation energy. Considering
the elastic energy associated with the infinite-dilution
limit, one obtains $E\p=$ -1.872 eV for 1 f.u. of  \ce{Mg5Si6},
$E\p=$ -1.486 eV/f.u. for \ce{Mg5Al2Si4}, and 
$E\p=$ -1.278 eV/f.u. for \ce{Mg4Al3Si4}. The formation
of the $\beta''$ phases starting from the SSS is so 
exoenergetic that needle-like precipitates can form
without overcoming a free energy barrier.  Due to the much lower 
surface energy for the C interface, in the small-precipitate limit \ce{Mg5Si6} forms the most stable precipitate. 
In the limit of macroscopic precipitates, the energy per f.u. tends to the precipitation energy plus the dilute-limit elastic contribution, given as 
$E_\infty=$ -2.467 eV/f.u. for  \ce{Mg5Si6},
$E_\infty=$ -2.651 eV/f.u. for \ce{Mg5Al2Si4}, and 
$E_\infty=$ -2.292 eV/f.u. for \ce{Mg4Al3Si4}. Thus, \ce{Mg5Al2Si4} is predicted to be 
the most stable form in the large-precipitate limit.  The elastic strain energy does not change the order of stability
but does narrow the energy difference between the most and least stable down to ~0.35 eV/f.u. or 0.032 eV/atom

\subsection{DFT of needle-shaped precipitates and comparison to CNT model}

The CNT model of precipitate energetics we have introduced in Eq.~\ref{eq:tdmodel}, including self-consistent elasticity terms, could be 
very useful to examine the interaction between growing precipitates. 
In order to assess its accuracy, we use the same needle-like geometry to evaluate the energetics of
precipitates using DFT, and perform a comparison with the results of the model.
To be consistent with the definition of formation energies used
above, we define the precipitation energy using the SSSS as reference, i.e. 
\begin{equation}
E\p(N)= E_{sys}^{tot}(N)-M\,E^{ss}_{\ce{Al}}
-N\sum_{x=\ce{Si, Mg, Al}} n_{x}\cdot E^{ss}_{x},
\label{eq:commonprecipitation}
\end{equation}
where $M$ is the %
number of \ce{Al} atoms in the matrix for a give simulation supercell,  and  $n_{x}$ and $E^{ss}_{x}$ indicate 
the $\beta''$ composition and the
solid-solution energy for \ce{Al},  \ce{Si} and \ce{Mg}, 
as in Eq.~\eqref{eq:formation}.

To benchmark the model across different precipitate sizes, 
we study three systems whose
cross-section contains
1, 4, and 16 formula units of precipitate in an equiaxed geometry.  
These precipitates are embedded in an Al matrix supercells of sizes ($a\times b\times c$)
 $5\times 1\times 5$, 
$7\times 1\times 7$, and $12\times 1\times 12$ \emph{fcc} unit cells, respectively,
as shown in Fig. \ref{fig:2dprecipitates}(a) 
for the supercell containing 16 f.u. of the $\beta''$ phase).

As noted above, the elastic energy depends on the precipitate density or cell geometry.
The DFT cells are not in the dilute limit.  Therefore, for comparison to the DFT
energies, the CNT model is modified to account for the elastic energy changes in the non-dilute 
limit as 
\begin{equation}
E\p(N) = N E_{form} + N E_{strain}(N,V) + E_{surf}(N), 
\label{eq:tdmodel1}
\end{equation}
where $E_{strain}(N,V)$ is the elastic strain energy per formula unit in a supercell
of volume V containing a precipitate of size $N$ formula units.  Elasticity calculations
have been performed using the method described earlier for precisely the geometries
studied in DFT, and the strain energies $E_{strain}(N,V)$ are shown in Table \ref{tab:valori}.  These
values are generally larger than the dilute limit, and increase with increasing $N$ due the larger fraction of $\beta''$ precipitate 
included in the supercell.

\begin{figure}[th!]
\centering  \includegraphics[width=0.85\columnwidth]{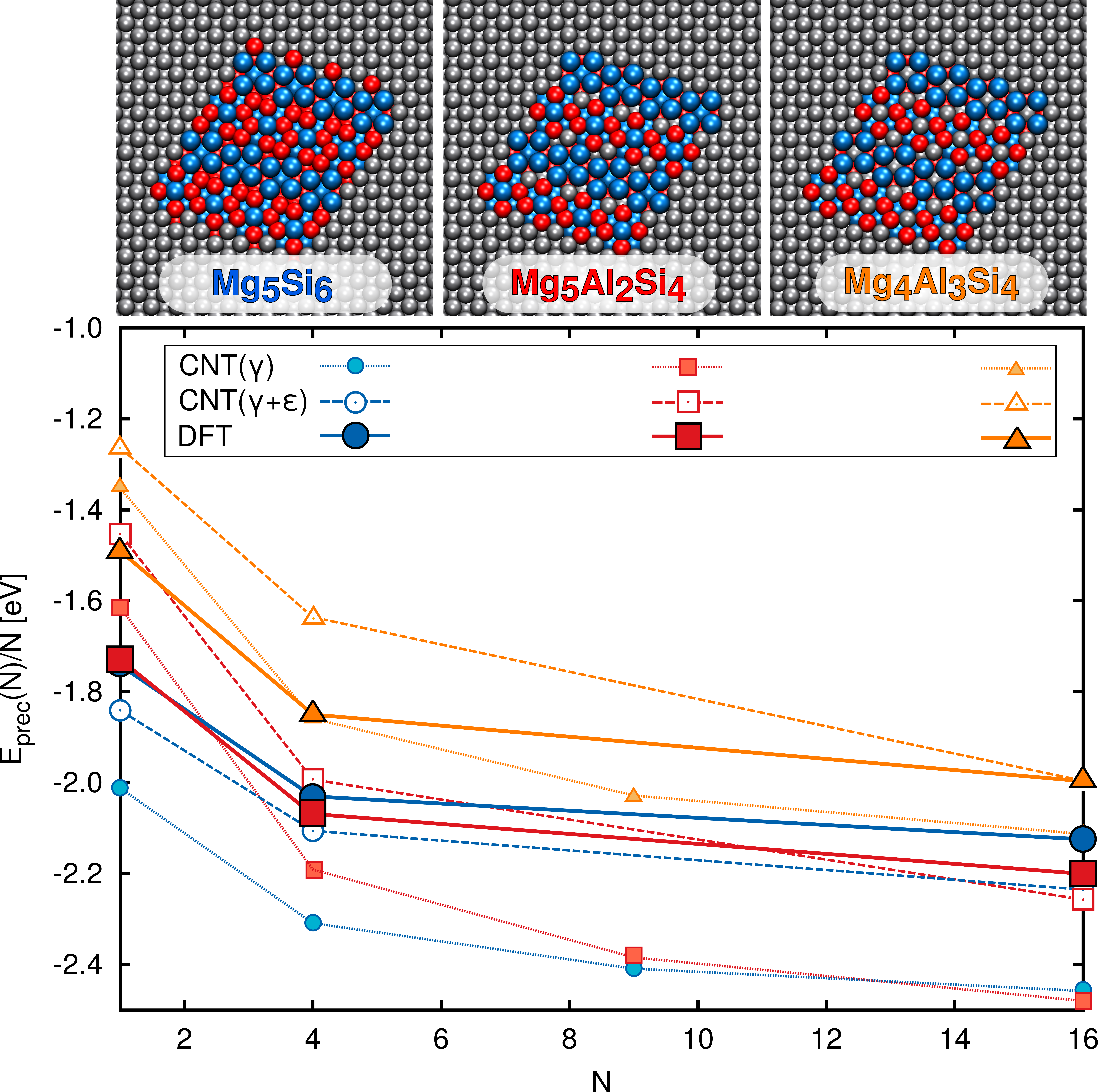}
\caption{\label{fig:precdft+tdsfs} (Top) A view along the $\mathbf{b}$ 
direction of an infinite needle-shaped $\beta''$ precipitate with 
16 (4x4) formula units cross-section, embedded in a $12\times 1\times 12$ Al supercell for the three precipitate
compositions.
(Bottom) Precipitation energies per
formula unit, $E_{prec}(N)/N$, 
calculated from explicit DFT calculations 
using Eqn.~\ref{eq:commonprecipitation} and estimated
from the thermodynamic CNT model in Eq. \ref{eq:tdmodel}.} 
\end{figure}

Figure~\ref{fig:precdft+tdsfs} compares the DFT precipitate energies, per formula unit, versus precipitate
size with predictions obtained using (i) surface energy terms only (CNT($\gamma$)) and 
(ii)  surface energies terms plus elastic strain energy in the DFT simulation cell (CNT($\gamma+\epsilon$)).  
The results generally follow the expected trend, in that larger precipitates are thermodynamically
more stable due to the reduction in relative importance of the interface, edge, and corner energies with increasing size, and
the energies approach the (size-independent) formation energies plus dilute-limit elastic energies for each of the 
three stoichiometries (Table~\ref{tab:valori}).
A CNT model that uses only the surface energies captures qualitatively the
asymptotic behavior for different $\beta''$ compositions, as
well as relative ordering. However, it under-estimates 
the energy of the precipitates in the large-precipitate-size limit, 
due to the absence of the positive contribution of the elastic energy.

The CNT($\gamma+\epsilon$) model predicts quite accurately the energetics of the
larger precipitates.  However, it significantly overestimates the energy at
the smaller sizes.  The full DFT energies are up 
to 0.4eV/f.u \emph{lower} than predicted by 
Eqn~\eqref{eq:tdmodel1}. One would normally expect that 
edge and corner terms would destabilize the nucleus (increase the energy) at the smaller
sizes.  Thus, the fact that the self-consistent energetics leads to stronger
stabilization suggests that the surface energies computed
assuming ideal interfaces provides
only an upper-bound to the actual $\gamma^{\text{A,B,C}}$.  
Further relaxation (which is hindered for the larger precipitates, and for periodic
surface calculations) 
could significantly lower the interface energy. 
Searching for reconstructions of the $\beta''\parallel$Al interfaces 
with a top-down approach and using electronic structure calculations
constitutes a formidable challenge.  We expect that the development of
machine-learning models\cite{Ryo} for classical inter-atomic potentials, 
together with Monte Carlo sampling techniques, 
might help elucidate this important
contribution to the stability and morphology of precipitates in the
Al-6000 series.

\begin{figure}[th!]
\centering  \includegraphics[width=0.85\columnwidth]{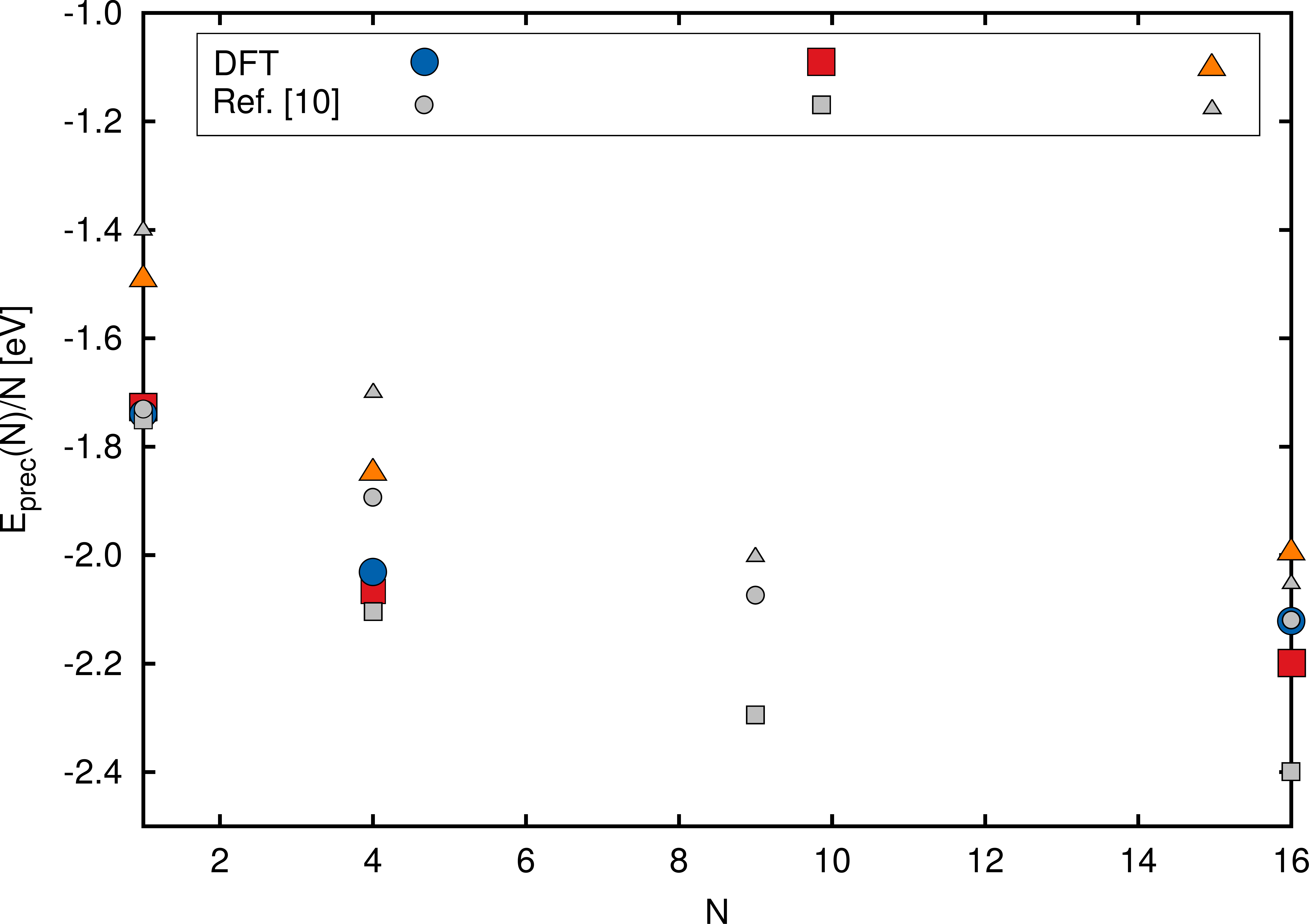}
\caption{\label{fig:2dprecipitates} 
Precipitation energies per
formula unit, $E_{prec}(N)/N$, 
calculated from explicit DFT calculations 
using Eqn.~\ref{eq:commonprecipitation}. 
Results from literature 
calculations that apply the same equation but include relaxation of the supercell are shown for comparison~\cite{niniveIIpaperthesis2014}. Symbols and colors are consistent with Fig. \ref{fig:precdft+tdsfs}.}
\end{figure}

Comparison between the calculations we report here and  those presented in
Ref.~\citenum{niniveIpaperthesis2014}
underscore the importance of accounting for elastic relaxation in this kind of simulations. While part of the discrepancy could be attributed to minor differences in the
computational details, we note a general trend where
the energies for the $4\times 4$ precipitates reported by Ref.~\citenum{niniveIpaperthesis2014} 
are considerably lower than those for the smaller $2\times 2$ precipitates, 
values -- and in all cases but for \ce{Mg5Si6} -- lower than our values.
As shown in the Appendix, this trend can be understood in terms of the boundary conditions chosen for DFT calculations.
Simulations in Ref.~\citenum{niniveIpaperthesis2014} allowed the
supercell dimensions to relax, which underestimates the energy of the encapsulated precipitate relative to the dilute limit. In our calculations, instead, we fixed 
fixed the cell parameters to match the Al bulk lattice parameter, which, conversely, overestimates the energy. 
Use of a fixed supercell simplifies the comparison between calculations, and the definition of consistent surface energies. However, only a multi-scale analysis that includes a FE model makes it possible to compute the elastic corrections to the ``dilute'' limit and to interpret quantitatively DFT results in terms of the physical contributions to the precipitate energy.

\section{Nucleation of a precipitate in 3D}
\label{sec:threed}

The analyses in the previous section show that there is no barrier for the growth of needle-like precipitates
starting at the smallest size $N=1$ for the in-plane precipitate structure.  The 
inclusion of interface and elastic energies was essential in this analysis 
to verify that nanoscopic precipitates are stable despite the high interface and elastic energy contributions.
We note that possible lower-energy interfaces will only enhance the stabilization of the 
smallest precipitates.  Therefore, nucleation of all three $\beta''$
phases studied here occurs at the in-plane unit cell level or below.
However, the in-plane analysis neglects the additional energy cost of the high-energy
$B$ $[010]_{\beta''}$ interface.  We thus investigate here the formation energy of 3D precipitates, to better understand the 
precipitate nucleation process and possible nucleation barriers.

We simulated 3D precipitates composed of a single formula unit
fully-embedded in the Al matrix.
As shown in Table~\ref{fig:3dprecipitates},  the fully-relaxed
DFT energy is negative for all compositions.  This confirms that 
precipitation is barrierless down to a single 3D formula unit even
when considering the high-$\gamma$ $B$ interfaces.
At this scale, the CNT($\gamma$) model is very inaccurate, predicting
positive formation energy for all the 
stoichiometries except \ce{Mg5Si6}.  The elastic strain energy computation
requires a full 3d analysis, and is not performed here since the elastic term would
increase the energy relative to the CNT($\gamma$) model.

It is not surprising that a mesoscopic model
cannot capture the energetics of a precipitate 
that consists of just eleven atoms. It is however 
interesting that -- just as for the needle-like 
geometry -- the mesoscale model \emph{overestimates} the energy cost associated with the precipitate-matrix interfaces, indicating that local relaxations can significantly lower the interface excess energy as compared to the ideal unreconstructed interfaces. 

\begin{table}[th!]
\begin{center}
\begin{tabular}[b]{   c  c c c   }     
   & \includegraphics[width=0.23\columnwidth]{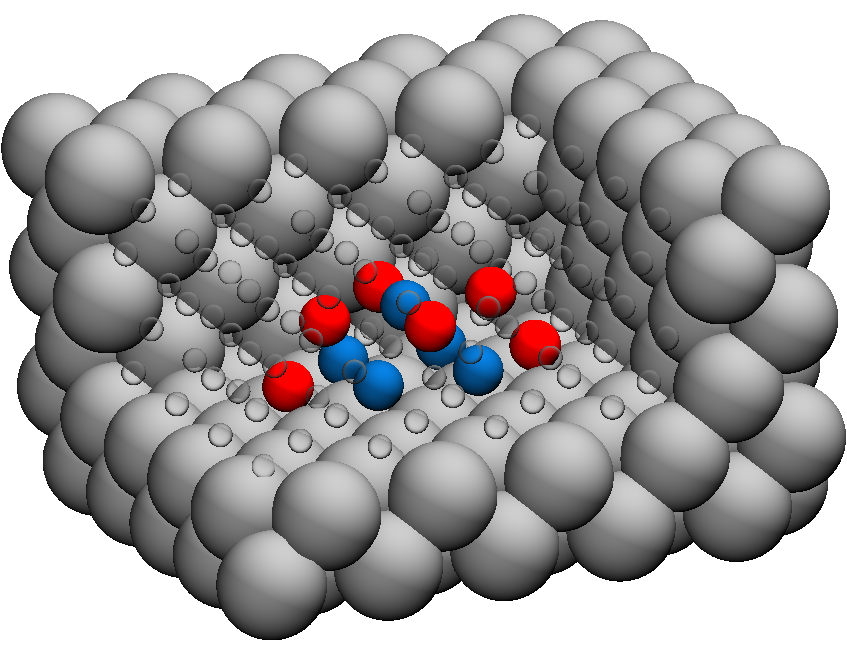}  & \includegraphics[width=0.23\columnwidth]{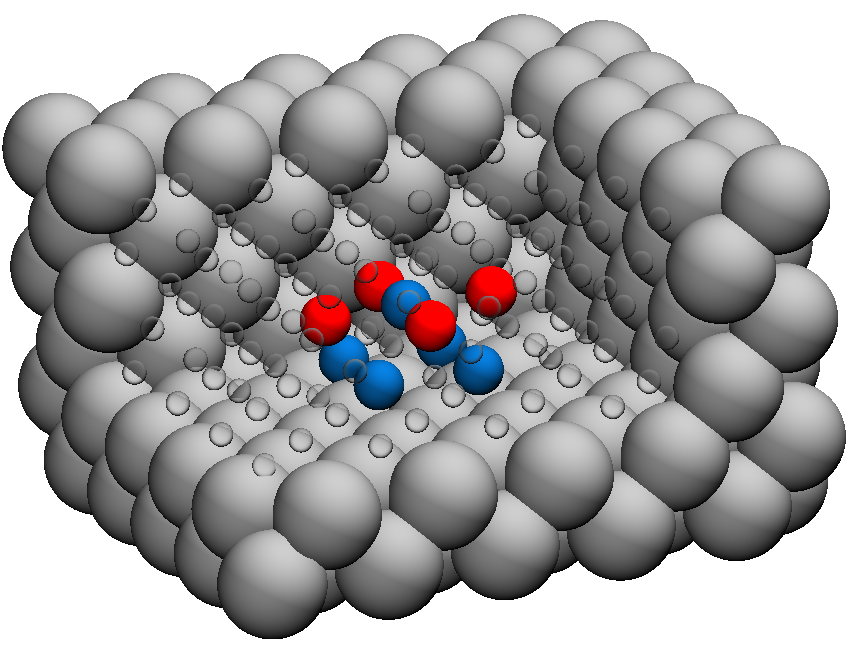} & \includegraphics[width=0.23\columnwidth]{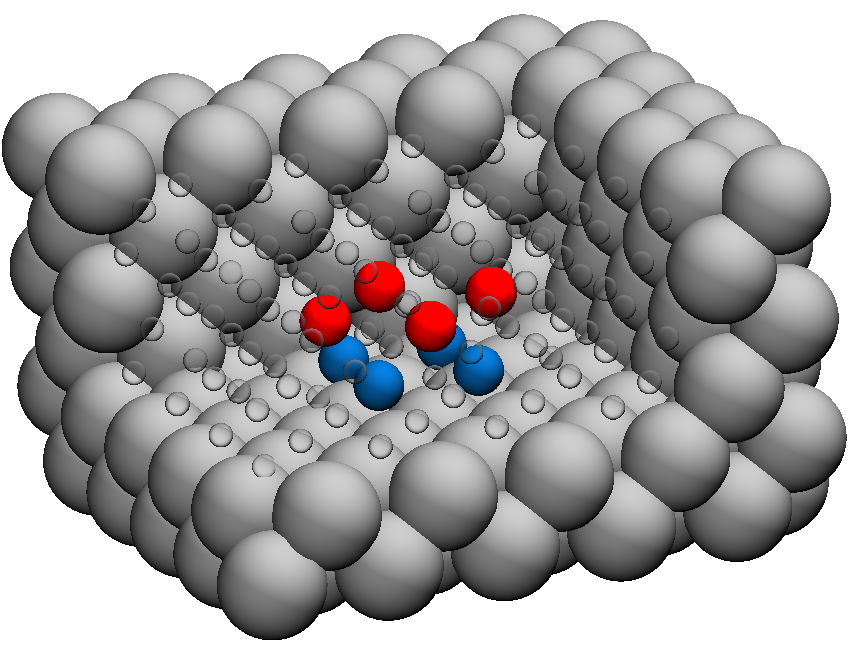}  \\  
\hline\hline\multicolumn{1}{c|}{$E_{prec}$ [meV]} & $Mg_5Si_6$  & $Mg_5Al_2Si_4$ & \multicolumn{1}{c}{$Mg_4Al_3Si_4$}  \\  
\hline  \multicolumn{1}{c|}{DFT}   & -658   & -558  &-351  \\  
  \multicolumn{1}{c|}{CNT($\gamma$)}   & -102 %
  & 513 %
  & 497 \\
\hline\hline  \end{tabular}

\caption{\label{fig:3dprecipitates}{(Top)} Snapshots of the simulation cells used to model single-formula-unit precipitates fully-encapsulated in the Al-matrix.  {(Table)} Precipitation energies of one formula unit precipitates computed from DFT 
calculations (Eqn.~\eqref{eq:commonprecipitation}) and the CNT model~\eqref{eq:tdmodel} with and without the finite element strain term. The area of the different interfaces has been assumed to correspond to those of half of the monoclinic unit cell.}
\end{center}
\end{table}

\section{Conclusion}

By clearly identifying the chemical, surface, and elastic strain energies
that contribute to the total precipitation energy versus size and composition, 
and demonstrating that the overall trends are consistent
with a thermodynamic classical-nucleation-theory-like
model, we have provided new insights into the early stages of the formation
of $\beta''$ precipitates in Al-6000 alloys.

The in-situ needle-like $\beta"$ precipitates are found to be stable relative
to the solid solution down to the smallest in-plane formula unit, indicating
barrier-less growth at and above this size.  The composition dependence of the total 
energies is subtle,
with two compositions being quite close in energy.  Thus, the inclusion of
surface energies and elastic energies due to the different precipitate structures and
compositions is essential for interpreting the DFT results and for then 
determining the energetics in the more-dilute limit of real materials.  
The benchmarking of the CNT-type model also provides a validation for the
use of such mesoscopic models in other systems.

The largest discrepancy between the thermodynamic CNT model and DFT 
calculations is seen for the smallest precipitates, with the \emph{ab initio} energies 
being consistently much \emph{lower} 
than those predicted based on surface energies
computed for a coherent interface between the precipitate and
the matrix. Together with the fact that the anisotropy of $\gamma$
is not sufficient to justify the aspect ratio of needle-like
$\beta''$ precipitates, this observation hints strongly at 
the need for consideration of more complex models of the interfaces
of the precipitates -- including variable composition and 
a significant degree of reconstruction -- that may help reduce
the interface and elastic energies and further stabilize the 
small precipitates.

We further show that, down to a single formula unit that is fully
encapsulated in the Al matrix, the DFT energy of a nanoscale
precipitate is lower than the reference supersaturated solid 
solution.  This underscores the fact that precipitation
kinetics is likely to be diffusion-limited. 
Aggregates of a few solute atoms that can act as vacancy traps~\cite{Pogatscher2011} would thus slow vacancy-mediated solute diffusion that is necessary to form larger
precipitates, greatly affecting the aging times.  This conclusion of dominance of diffusion-controlled aging
is also consistent with recent findings that the addition of 100 ppm of Sn to Al-6061 can significantly delay
aging, attributed to trapping of the quenched-in vacancies by the Sn atoms~\cite{poga+14prl,Francis2016}.
Our results thus point toward the need for a systematic study of the energetics of
aggregates in the GP-zone regime, and the interactions between
those aggregates and vacancies and/or trace elements in the alloy
to understand and fine-tune the
behavior of Al-6000 alloys in the early stages of precipitation.

\paragraph{Acknowledgements} The authors acknowledge 
insightful discussion with
Dr.  Christophe Sigli and 
Dr. Timothy Warner.
DG and MC acknoweledge support for this work by an Industial Research Grant
funded by Constellium. TJ and WC acknowledge support for this work 
through a European Research Council Advanced Grant, 
``Predictive Computational Metallurgy", ERC Grant agreement No. 339081 - PreCoMet.

\appendix

\section{Calculation of eigenstrain and stiffness tensors}
\label{sec:eigenstrain}
The eigenstrain $\bar{\boldsymbol{\epsilon}}$ is the strain required to compensate for the misfit between the matrix and precipitate lattices, i.e., the strain that deforms a formula unit of precipitate into the shape of a formula unit of undeformed matrix.
Subsequently, we show how to compute $\bar{\boldsymbol{\epsilon}}$ in the global frame of reference $\hat{\vec e}_x$-$\hat{\vec e}_y$-$\hat{\vec e}_z$ described in Figure~\ref{fig:orientation}. The formula unit geometries of the matrix and the precipitates are monoclinic cells for which the directions of $\vec c$ and $\vec b$ coincide but differ in the angle $\beta$ and the edge lengths $a, b, c$. We start by determining the material frame of reference $\hat{\vec e}_\alpha$-$\hat{\vec e}_\beta$-$\hat{\vec e}_z$ as it  simplifies both the expression of the edge vectors $\vec a, \vec b, \vec c$ and, since the elastic constants reported in Table~\ref{tab:elastic_constants} are computed in that frame, is required to compute stiffness tensors in the global frame.

\begin{figure}[htb]
  \centering
  \includegraphics{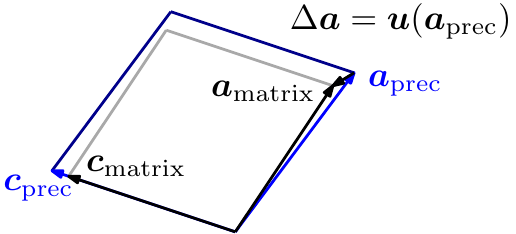}
\quad \includegraphics[height=1in]{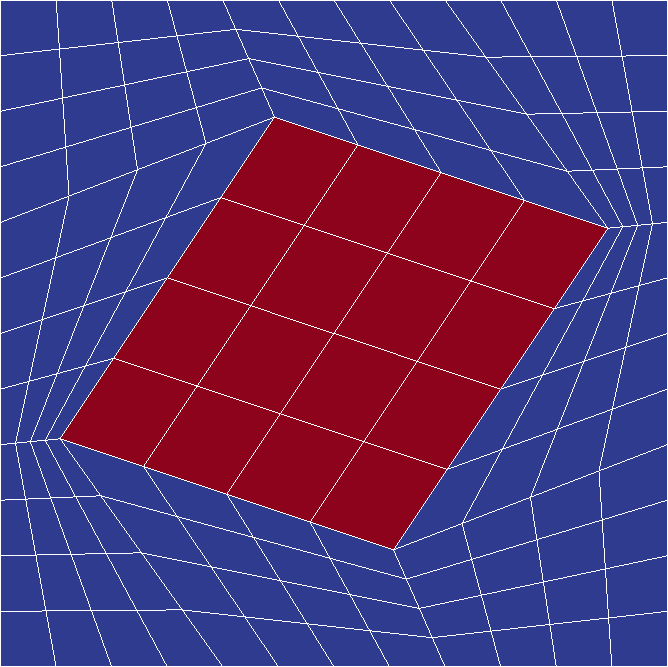}\quad
\includegraphics[height=1in]{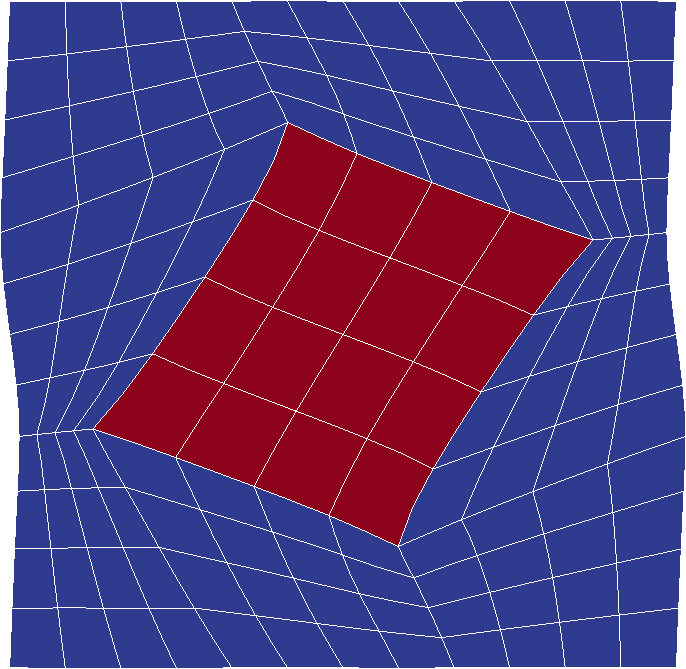}
  \caption{Schematic illustration of eigendisplacement (left). Mesh for finite element analysis of the elastic problem (center). Note that the structured mesh follows the boundary of the precipitate (red parallelogram). Deformed elastic problem (right). The displacements have been magnified by 5 for better visibility. Note the periodic deformation.}
  \label{fig:strain_calc}
\end{figure}

The basis vectors $\hat{\vec e}_\beta$ and $\hat{\vec e}_z$ are collinear with the formula unit cell edge vectors defined in \eqref{eq:interfacerel}, $\vec c$ and $\vec b$, respectively, and the third basis vector $\hat{\vec e}_\alpha$ is chosen to complete a right-handed orthonormal basis
\begin{equation}
  \label{eq:material-basis}
  \hat{\vec e}_\beta  = \frac{\vec c}{c} = \tfrac{1}{\sqrt{10}}\left(-3, 1, 0\right)^\text{T},\ 
  \hat{\vec e}_z = \frac{\vec b}{b} = \left(0, 0, 1\right)^\text{T},\ 
  \hat{\vec e}_\alpha = \hat{\vec e}_\beta\times\hat{\vec e}_z =  \tfrac{1}{\sqrt{10}}\left(1, 3, 0\right)^\text{T}.
\end{equation}
We use the basis vectors to express the edge vectors in the global frame of reference using Table~\ref{tab:lattice}
\begin{equation}
  \label{eq:edge_vects}
  \vec c = c\,\hat{\vec e}_\beta,\ \vec b = b\,\hat{\vec e}_z,\ \vec a = a\,\left(\sin{\beta}\,\hat{\vec e}_\alpha + \cos{\beta}\,\hat{\vec e}_\beta\right).
\end{equation}
The eigenstrain $\bar{\boldsymbol{\epsilon}}$ corresponds to a displacement gradient $\boldsymbol \nabla\vec u$ that transforms the precipitate edge vectors into the matrix edge vectors, see Figure~\ref{fig:strain_calc} (left).
After defining matrices composed of the edge vectors for a precipitate $\ten V\p = \left(\vec a\p, \vec b\p,\vec c\p\right)$ and the matrix $\ten V\m = \left(\vec a\m, \vec b\m,\vec c\m\right)$, the displacement gradient $\boldsymbol \nabla\vec u$ can be expressed as

\begin{equation}
  \label{eq:def_grad}
  \ten V\m = \boldsymbol\nabla\vec u\,\ten V\p + \ten V\p\ \Rightarrow\ \boldsymbol\nabla \vec u= \ten V\m\,\ten V\p^{-1}-\ten I,
\end{equation}
where $\ten I$ is the identity matrix. The eigenstrain $\bar{\boldsymbol{\epsilon}}$ is the symmetric part of $\boldsymbol\nabla\vec u$
\begin{equation}
  \label{eq:eigenstrain}
  \bar{\boldsymbol{\epsilon}} = \tfrac12\left(\boldsymbol\nabla\vec u + \boldsymbol\nabla\vec u^\text{T}\right).
\end{equation}
The elastic constants of the precipitates have been calculated in the material frame of reference $\hat{\vec e}_\alpha$-$\hat{\vec e}_\beta$-$\hat{\vec e}_z$ and the corresponding stiffness tensor has to be rotated into the global frame of reference for the finite-element analysis. The stress $\boldsymbol\sigma$ and strain $\boldsymbol\epsilon$ in the global frame of reference are related to the material frame stress $\boldsymbol\sigma'$ and strain $\boldsymbol\epsilon'$ by the rotation $\ten R = \left(\hat{\vec e}_\alpha,\hat{\vec e}_\beta,\hat{\vec e}_z\right)$
\begin{equation}
  \label{eq:basechange}
  \boldsymbol \epsilon' = \ten R\boldsymbol \epsilon\ten R^\text{T},\quad \boldsymbol \sigma' = \ten R\boldsymbol \sigma\ten R^\text{T},
\end{equation}
and the relationship between $\boldsymbol\sigma'$ and $\boldsymbol\epsilon'$ is governed by elasticity
\begin{equation}
  \label{eq:elasticity}
  \boldsymbol\sigma' = \ten C'\,\boldsymbol\epsilon',
\end{equation}
where $\ten C'$ is the stiffness tensor in the material frame of reference. The stiffness tensor in the global frame of reference $\ten C$ can be obtained by combination \eqref{eq:basechange} and \eqref{eq:elasticity} in index notation (Einstein summation applies to repeated indices)
\begin{align}
  R_{ij}\sigma_{jk}R_{lk} &= C'_{ilmn}R_{mo}\epsilon_{op}R_{np},\notag\\
  \underbrace{R_{ia}R_{ij}}_{\delta_{aj}}\sigma_{jk}\underbrace{R_{lk}R_{lb}}_{\delta_{kb}} &=
  R_{ia}C'_{ilmn}R_{mo}\epsilon_{op}R_{np}R_{lb},\notag\\
  \sigma_{ab} &= R_{ia}R_{lb} C'_{ilmn}R_{mo}R_{np}\epsilon_{op},\notag\\
                C_{abop} &= R_{ia}R_{lb}R_{mo}R_{np}C'_{ilmn}.
\end{align}

\section{Elastic calculations}
\label{sec:elastic_calculations}

The elastic calculations use the finite element method \citep{zienkiewicz_finite_1977} and have been performed using a modified version of the open-source finite-element code Akantu \citep{richart2015implementation}. This section explains the chosen procedure.

We modeled the elastic problem using a structured, quadrilateral, and periodic two-dimensional mesh of bi-quadratic serendipity elements with eight nodes \citep{ergatoudis1968isoparametric}. The element type was chosen over linear elements for its high accuracy in static problems. In order to enforce periodic boundary conditions, we define the boundary nodes $i_s$ of the upper and right boundary as slave nodes to their counterparts on the bottom and left boundary (master nodes $i_m$). During the evaluation of nodal forces on master nodes $\vec f_{i_m}$, the forces acting their slave nodes are also assembled on the master $\vec f_{i_m}^\text{tot} = \vec f_{i_m} + \vec f_{i_s}$ and the slave node displacement is set to be equal to the displacement of their master $\vec u_{i_s} = \vec u_{i_m}$. In order to preclude solid body motion (and, thus, a singular stiffness matrix $\ten K$), the center node in the precipitate is fully blocked $\vec u_\text{c} = \boldsymbol 0$.

\begin{figure}[htb]
  \centering
  \includegraphics[width=\textwidth]{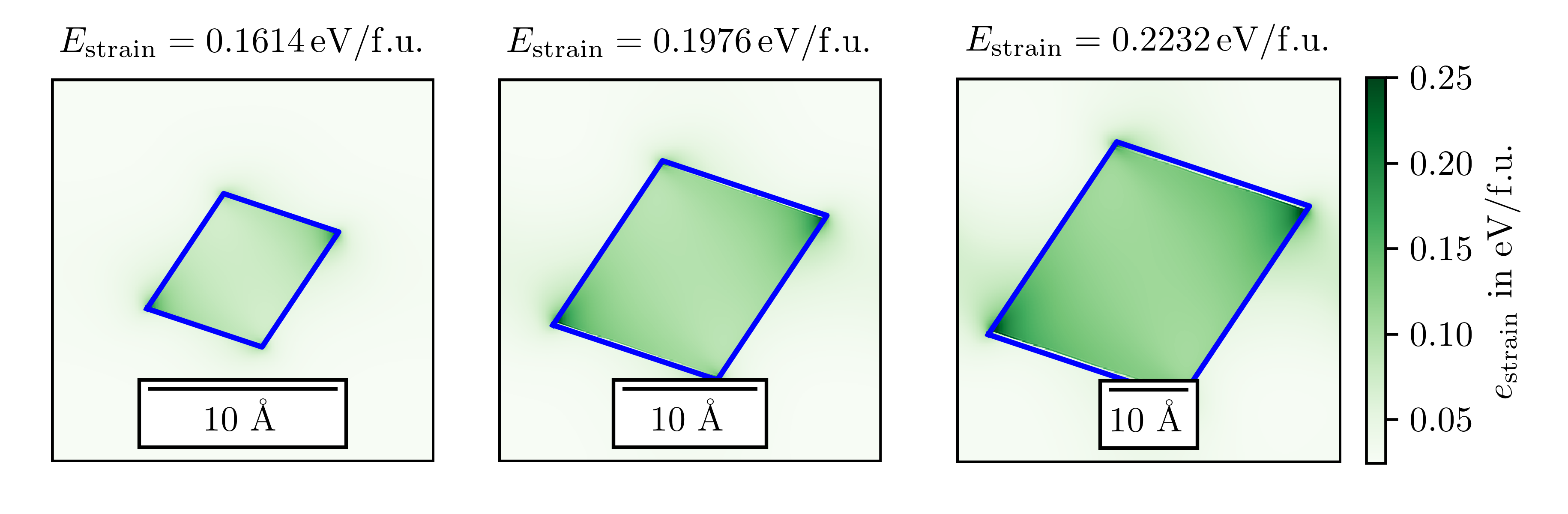}
  \caption{Distribution of strain energy density $e_\text{strain}$ for different geometries in the example of \ce{Mg4Al3Si4}. The blue frame marks the boundaries of the precipitate.}
  \label{fig:strain_energy}
\end{figure}

Figure \ref{fig:strain_calc} (center and right) shows such a mesh in its original and deformed state where the displacements have been amplified by a factor five for better visibility. The structured mesh follows the boundary of the precipitate, such that any element is either of matrix material (blue) or precipitate material(red). Note the periodic deformation of the simulation cell. The precipitate is preloaded with the eigenstrain $\bar{\boldsymbol\epsilon}$ as described in Section~\ref{sec:eigenstrain} and the stiffness tensors for matrix $\ten C\m$ and precipitate and $\ten C\p$ are assigned to the blue and red elements respectively. In absence of external loads, the assembled system of equations to solve is
\begin{equation}
  \label{eq:fem_system}
  \ten K\, \vec U = \vec 0,
\end{equation}
where $\ten K$ is the assembled stiffness matrix and $\vec U$ is the vector of all displacement degrees of freedom. We solve this system using the direct solver Mumps \citep{amestoy98mumpsmultifrontal}. The calculation of strain energy exploits the quadrature routines of Akantu using the shape functions of the elements to evaluate the integrals in \eqref{eq:strain_energy}. Figure~\ref{fig:strain_energy} shows the distribution of strain energy density $e_\text{strain}$ for the geometries considered using the example of \ce{Mg4Al3Si4}.
A mesh that is eight times finer than the one represented in Figure~\ref{fig:strain_calc} was used for smooth visualization.

\newcommand{\dV}{\mathrm dV}
\subsection{Relaxation of boundary conditions}
In order to compare our results more readily to those presented in \cite{niniveIpaperthesis2014}, we have additionally performed elastic calculations with fully relaxed periodic boundary conditions, in which the simulation box was allowed to expand and tilt as needed to have no average stress. This was done by following the procedure described in Appendix~\ref{sec:elastic_calculations}, but with an additional uniform eigenstrain added to all elements. This additional eigenstrain was used as a degree of freedom in a minimization of the total strain energy.

Table~\ref{tab:relax_bnd_cond} compares the strain energies per formula unit obtained with fixed periodic boundary conditions like the ones used in all DFT calculations in this work to the energies obtained using the relaxed boundary conditions used in \cite{niniveIpaperthesis2014}. One can see that the relaxed conditions lead to a consistent underestimation of the strain energy, while the fixed periodic conditions lead to overestimated energies.
\begin{table}[h]
  \centering
  \begin{tabular}{l|r|r r|r r|r r}
  \hline\hline
    Composition  & dilute& \multicolumn{2}{c|}{$4\times 4$} & \multicolumn{2}{c|}{$2\times 2$}  & \multicolumn{2}{c}{$1\times 1$} \\
    \text{[meV per f.u.]}                  &       & \multicolumn{1}{c}{fixed} & \multicolumn{1}{|c}{relaxed}  & \multicolumn{1}{|c}{fixed} & \multicolumn{1}{|c}{relaxed}  & \multicolumn{1}{|c}{fixed} & \multicolumn{1}{|c}{relaxed}  \\\hline
    \ce{Mg4Al3Si4} & 74 &117& 53&106& 59& 89& 66 \\
    \ce{Mg5Al2Si4} &128 &223& 89&198& 98&161&113 \\
    \ce{   Mg5Si6} &140 &223&114&203&122&171&132\\
    \hline\hline
  \end{tabular}

\caption{Comparison of elastic strain energies $E_{strain}$ obtained for all considered geometries with periodic boundary conditions of fixed dimensions (as the DFT calculations in this work) or fully relaxed conditions for which there is no mean stress on the simulation box (as in \cite{niniveIpaperthesis2014}).}
\label{tab:relax_bnd_cond}
\end{table}


\section*{References}

\begin{thebibliography}{10}
\expandafter\ifx\csname url\endcsname\relax
  \def\url#1{\texttt{#1}}\fi
\expandafter\ifx\csname urlprefix\endcsname\relax\def\urlprefix{URL }\fi
\expandafter\ifx\csname href\endcsname\relax
  \def\href#1#2{#2} \def\path#1{#1}\fi

\bibitem{murayama_pre-precipitate_1999}
M.~Murayama, K.~Hono,
  \href{http://linkinghub.elsevier.com/retrieve/pii/S1359645499000336}{Pre-precipitate
  clusters and precipitation processes in {Al}–{Mg}–{Si} alloys}, Acta
  Materialia 47~(5) (1999) 1537--1548.
\newblock \href {http://dx.doi.org/10.1016/S1359-6454(99)00033-6}
  {\path{doi:10.1016/S1359-6454(99)00033-6}}.
\newline\urlprefix\url{http://linkinghub.elsevier.com/retrieve/pii/S1359645499000336}

\bibitem{edwards_precipitation_1998}
G.~Edwards, K.~Stiller, G.~Dunlop, M.~Couper,
  \href{http://linkinghub.elsevier.com/retrieve/pii/S1359645498000597}{The
  precipitation sequence in {Al}–{Mg}–{Si} alloys}, Acta Materialia 46~(11)
  (1998) 3893--3904.
\newblock \href {http://dx.doi.org/10.1016/S1359-6454(98)00059-7}
  {\path{doi:10.1016/S1359-6454(98)00059-7}}.
\newline\urlprefix\url{http://linkinghub.elsevier.com/retrieve/pii/S1359645498000597}

\bibitem{ringer_microstructural_2000}
S.~Ringer, K.~Hono,
  \href{http://linkinghub.elsevier.com/retrieve/pii/S1044580399000510}{Microstructural
  {Evolution} and {Age} {Hardening} in {Aluminium} {Alloys}}, Materials
  Characterization 44~(1-2) (2000) 101--131.
\newblock \href {http://dx.doi.org/10.1016/S1044-5803(99)00051-0}
  {\path{doi:10.1016/S1044-5803(99)00051-0}}.
\newline\urlprefix\url{http://linkinghub.elsevier.com/retrieve/pii/S1044580399000510}

\bibitem{ravi-wolv04am}
C.~Ravi, C.~Wolverton, {First-principles study of crystal structure and
  stability of Al/Mg/Si/(Cu) precipitates}, Acta Materialia 52 (2004)
  4213--4227.

\bibitem{marioara_influence_2005}
C.~D. Marioara, S.~J. Andersen, H.~W. Zandbergen, R.~Holmestad,
  \href{http://link.springer.com/article/10.1007/s11661-005-0185-1}{The
  influence of alloy composition on precipitates of the {Al}-{Mg}-{Si} system},
  Metallurgical and Materials Transactions A 36~(3) (2005) 691--702.
\newline\urlprefix\url{http://link.springer.com/article/10.1007/s11661-005-0185-1}

\bibitem{takeda_stability_1998}
M.~Takeda, F.~Ohkubo, T.~Shirai, K.~Fukui,
  \href{http://link.springer.com/article/10.1023/A:1004355824857}{Stability of
  metastable phases and microstructures in the ageing process of
  {Al}–{Mg}–{Si} ternary alloys}, Journal of materials science 33~(9)
  (1998) 2385--2390.
\newline\urlprefix\url{http://link.springer.com/article/10.1023/A:1004355824857}

\bibitem{andersen_crystal_1998}
S.~J. Andersen, H.~W. Zandbergen, J.~Jansen, C.~Traeholt, U.~Tundal, O.~Reiso,
  \href{http://www.sciencedirect.com/science/article/pii/S135964549700493X}{The
  crystal structure of the β ″phase in {Al}–{Mg}–{Si} alloys}, Acta
  Materialia 46~(9) (1998) 3283--3298.
\newline\urlprefix\url{http://www.sciencedirect.com/science/article/pii/S135964549700493X}

\bibitem{zandbergen_data_2015}
M.~W. Zandbergen, Q.~Xu, A.~Cerezo, G.~D.~W. Smith,
  \href{http://www.sciencedirect.com/science/article/pii/S2352340915002395}{Data
  analysis and other considerations concerning the study of precipitation in
  {Al}–{Mg}–{Si} alloys by {Atom} {Probe} {Tomography}}, Data in Brief 5
  (2015) 626--641.
\newblock \href {http://dx.doi.org/10.1016/j.dib.2015.09.045}
  {\path{doi:10.1016/j.dib.2015.09.045}}.
\newline\urlprefix\url{http://www.sciencedirect.com/science/article/pii/S2352340915002395}

\bibitem{marioara_influence_2003}
C.~Marioara, S.~Andersen, J.~Jansen, H.~Zandbergen,
  \href{http://linkinghub.elsevier.com/retrieve/pii/S1359645402004706}{The
  influence of temperature and storage time at {RT} on nucleation of the β″
  phase in a 6082 {Al}–{Mg}–{Si} alloy}, Acta Materialia 51~(3) (2003)
  789--796.
\newblock \href {http://dx.doi.org/10.1016/S1359-6454(02)00470-6}
  {\path{doi:10.1016/S1359-6454(02)00470-6}}.
\newline\urlprefix\url{http://linkinghub.elsevier.com/retrieve/pii/S1359645402004706}

\bibitem{niniveIpaperthesis2014}
P.~H. Ninive, A.~Strandlie, S.~Gulbrandsen-Dahl, W.~Lefebvre, C.~D. Marioara,
  S.~J. Andersen, J.~Friis, R.~Holmestad, O.~M. Løvvik, Detailed atomistic
  insight into the phase in al–mg–si alloys, Acta Materialia 69 (2014)
  126--134.
\newblock \href {http://dx.doi.org/10.1016/j.actamat.2014.01.052}
  {\path{doi:10.1016/j.actamat.2014.01.052}}.

\bibitem{poga+14prl}
S.~Pogatscher, H.~Antrekowitsch, M.~Werinos, F.~Moszner, S.~S.~A. Gerstl, M.~F.
  Francis, W.~A. Curtin, J.~F. L{\"{o}}ffler, P.~J. Uggowitzer, {Diffusion on
  Demand to Control Precipitation Aging: Application to Al-Mg-Si Alloys}, Phys.
  Rev. Lett. 112 (2014) 225701.

\bibitem{marioara_atomic_2001}
C.~D. Marioara, S.~J. Andersen, J.~Jansen, H.~W. Zandbergen,
  \href{http://www.sciencedirect.com/science/article/pii/S1359645400003025}{Atomic
  model for {GP}-zones in a 6082 {Al}–{Mg}–{Si} system}, Acta materialia
  49~(2) (2001) 321--328.
\newline\urlprefix\url{http://www.sciencedirect.com/science/article/pii/S1359645400003025}

\bibitem{derlet_first-principles_2002}
P.~M. Derlet, S.~J. Andersen, C.~D. Marioara, A.~Frøseth,
  \href{http://iopscience.iop.org/article/10.1088/0953-8984/14/15/315/meta}{A
  first-principles study of the β”-phase in {Al}-{Mg}-{Si} alloys}, Journal
  of Physics: Condensed Matter 14~(15) (2002) 4011.
\newline\urlprefix\url{http://iopscience.iop.org/article/10.1088/0953-8984/14/15/315/meta}

\bibitem{hasting_composition_2009}
H.~S. Hasting, A.~G. Fro̸seth, S.~J. Andersen, R.~Vissers, J.~C. Walmsley,
  C.~D. Marioara, F.~Danoix, W.~Lefebvre, R.~Holmestad,
  \href{http://scitation.aip.org/content/aip/journal/jap/106/12/10.1063/1.3269714}{Composition
  of β[sup ʺ] precipitates in {Al}–{Mg}–{Si} alloys by atom probe
  tomography and first principles calculations}, Journal of Applied Physics
  106~(12) (2009) 123527.
\newblock \href {http://dx.doi.org/10.1063/1.3269714}
  {\path{doi:10.1063/1.3269714}}.
\newline\urlprefix\url{http://scitation.aip.org/content/aip/journal/jap/106/12/10.1063/1.3269714}

\bibitem{gian+09jpcm}
P.~Giannozzi, S.~Baroni, N.~Bonini, M.~Calandra, R.~Car, C.~Cavazzoni,
  D.~Ceresoli, G.~L. Chiarotti, M.~Cococcioni, I.~Dabo, A.~D. Corso,
  S.~de~Gironcoli, S.~Fabris, G.~Fratesi, R.~Gebauer, U.~Gerstmann,
  C.~Gougoussis, A.~Kokalj, M.~Lazzeri, L.~Martin-Samos, N.~Marzari, F.~Mauri,
  R.~Mazzarello, S.~Paolini, A.~Pasquarello, L.~Paulatto, C.~Sbraccia,
  S.~Scandolo, G.~Sclauzero, A.~P. Seitsonen, A.~Smogunov, P.~Umari, R.~M.
  Wentzcovitch, {QUANTUM ESPRESSO: a modular and open-source software project
  for quantum simulations of materials}, J. Phys. Condens. Matter 21~(39)
  (2009) 395502--395519.

\bibitem{perd+96prl}
J.~P. Perdew, K.~Burke, M.~Ernzerhof, {Generalized Gradient Approximation made
  simple}, Phys. Rev. Lett. 77~(18) (1996) 3865.

\bibitem{vand90prb}
D.~Vanderbilt, {Soft self-consistent pseudopotentials in a generalized
  eigenvalue formalism}, Phys. Rev. B 41 (1990) 7892--7895.

\bibitem{kresse_ultrasoft_1999}
G.~Kresse, D.~Joubert,
  \href{http://journals.aps.org/prb/abstract/10.1103/PhysRevB.59.1758}{From
  ultrasoft pseudopotentials to the projector augmented-wave method}, Physical
  Review B 59~(3) (1999) 1758.
\newline\urlprefix\url{http://journals.aps.org/prb/abstract/10.1103/PhysRevB.59.1758}

\bibitem{materialcloud}
I.~Castelli, N.~Marzari, \href{http://materialscloud.org/}{Standard solid state
  pseudopotentials} (2015).
\newline\urlprefix\url{http://materialscloud.org/}

\bibitem{monk-pack76prb}
H.~J. Monkhorst, J.~D. Pack, {Special points for Brillouin-zone integrations},
  Phys. Rev. B 13~(12) (1976) 5188--5192.

\bibitem{davey_precision_1925}
W.~P. Davey,
  \href{http://journals.aps.org/pr/abstract/10.1103/PhysRev.25.753}{Precision
  measurements of the lattice constants of twelve common metals}, Physical
  Review 25~(6) (1925) 753.
\newline\urlprefix\url{http://journals.aps.org/pr/abstract/10.1103/PhysRev.25.753}

\bibitem{tambe_bulk_2008}
M.~J. Tambe, N.~Bonini, N.~Marzari,
  \href{http://link.aps.org/doi/10.1103/PhysRevB.77.172102}{Bulk aluminum at
  high pressure: {A} first-principles study}, Physical Review B 77~(17).
\newblock \href {http://dx.doi.org/10.1103/PhysRevB.77.172102}
  {\path{doi:10.1103/PhysRevB.77.172102}}.
\newline\urlprefix\url{http://link.aps.org/doi/10.1103/PhysRevB.77.172102}

\bibitem{nielsen_first-principles_1983}
O.~H. Nielsen, R.~M. Martin,
  \href{http://journals.aps.org/prl/abstract/10.1103/PhysRevLett.50.697}{First-principles
  calculation of stress}, Physical Review Letters 50~(9) (1983) 697.
\newline\urlprefix\url{http://journals.aps.org/prl/abstract/10.1103/PhysRevLett.50.697}

\bibitem{elasticonstant2002}
G.~V. Sin’ko, N.~A. Smirnov,
  \href{http://stacks.iop.org/0953-8984/14/i=29/a=301}{Ab initio calculations
  of elastic constants and thermodynamic properties of bcc, fcc, and hcp al
  crystals under pressure}, Journal of Physics: Condensed Matter 14~(29) (2002)
  6989.
\newline\urlprefix\url{http://stacks.iop.org/0953-8984/14/i=29/a=301}

\bibitem{bercegeay_first-principles_2005}
C.~Bercegeay, S.~Bernard,
  \href{http://link.aps.org/doi/10.1103/PhysRevB.72.214101}{First-principles
  equations of state and elastic properties of seven metals}, Physical Review B
  72~(21).
\newblock \href {http://dx.doi.org/10.1103/PhysRevB.72.214101}
  {\path{doi:10.1103/PhysRevB.72.214101}}.
\newline\urlprefix\url{http://link.aps.org/doi/10.1103/PhysRevB.72.214101}

\bibitem{yu_calculations_2010}
R.~Yu, J.~Zhu, H.~Ye,
  \href{http://linkinghub.elsevier.com/retrieve/pii/S0010465509003932}{Calculations
  of single-crystal elastic constants made simple}, Computer Physics
  Communications 181~(3) (2010) 671--675.
\newblock \href {http://dx.doi.org/10.1016/j.cpc.2009.11.017}
  {\path{doi:10.1016/j.cpc.2009.11.017}}.
\newline\urlprefix\url{http://linkinghub.elsevier.com/retrieve/pii/S0010465509003932}

\bibitem{li_computer_1998}
D.~Li, L.~Chen,
  \href{http://linkinghub.elsevier.com/retrieve/pii/S1359645497004783}{Computer
  simulation of stress-oriented nucleation and growth of θ′ precipitates
  {inAl}–{Cu} alloys}, Acta Materialia 46~(8) (1998) 2573--2585.
\newblock \href {http://dx.doi.org/10.1016/S1359-6454(97)00478-3}
  {\path{doi:10.1016/S1359-6454(97)00478-3}}.
\newline\urlprefix\url{http://linkinghub.elsevier.com/retrieve/pii/S1359645497004783}

\bibitem{luo_stress/strain_2014}
K.~Luo, B.~Zang, S.~Fu, Y.~Jiang, D.-q. Yi,
  \href{http://linkinghub.elsevier.com/retrieve/pii/S1003632614633239}{Stress/strain
  aging mechanisms in {Al} alloys from first principles}, Transactions of
  Nonferrous Metals Society of China 24~(7) (2014) 2130--2137.
\newblock \href {http://dx.doi.org/10.1016/S1003-6326(14)63323-9}
  {\path{doi:10.1016/S1003-6326(14)63323-9}}.
\newline\urlprefix\url{http://linkinghub.elsevier.com/retrieve/pii/S1003632614633239}

\bibitem{fu_effects_2014}
S.~Fu, D.-q. Yi, H.-q. Liu, Y.~Jiang, B.~Wang, Z.~Hu,
  \href{http://linkinghub.elsevier.com/retrieve/pii/S1003632614633458}{Effects
  of external stress aging on morphology and precipitation behavior of θ″
  phase in {Al}-{Cu} alloy}, Transactions of Nonferrous Metals Society of China
  24~(7) (2014) 2282--2288.
\newblock \href {http://dx.doi.org/10.1016/S1003-6326(14)63345-8}
  {\path{doi:10.1016/S1003-6326(14)63345-8}}.
\newline\urlprefix\url{http://linkinghub.elsevier.com/retrieve/pii/S1003632614633458}

\bibitem{yao_tem_2001}
J.-Y. Yao, D.~A. Graham, B.~Rinderer, M.~J. Couper,
  \href{http://www.sciencedirect.com/science/article/pii/S0968432800000950}{A
  {TEM} study of precipitation in {Al}–{Mg}–{Si} alloys}, Micron 32~(8)
  (2001) 865--870.
\newline\urlprefix\url{http://www.sciencedirect.com/science/article/pii/S0968432800000950}

\bibitem{wang_first-principles_2007}
Y.~Wang, Z.-K. Liu, L.-Q. Chen, C.~Wolverton,
  \href{http://linkinghub.elsevier.com/retrieve/pii/S1359645407004673}{First-principles
  calculations of β″-{Mg}5si6/α-{Al} interfaces}, Acta Materialia 55~(17)
  (2007) 5934--5947.
\newblock \href {http://dx.doi.org/10.1016/j.actamat.2007.06.045}
  {\path{doi:10.1016/j.actamat.2007.06.045}}.
\newline\urlprefix\url{http://linkinghub.elsevier.com/retrieve/pii/S1359645407004673}

\bibitem{niniveIIpaperthesis2014}
P.~H. Ninive, O.~M. Løvvik, A.~Strandlie,
  \href{http://link.springer.com/10.1007/s11661-014-2214-4}{Density
  {Functional} {Study} of the β″ {Phase} in {Al}-{Mg}-{Si} {Alloys}},
  Metallurgical and Materials Transactions A 45~(6) (2014) 2916--2924.
\newblock \href {http://dx.doi.org/10.1007/s11661-014-2214-4}
  {\path{doi:10.1007/s11661-014-2214-4}}.
\newline\urlprefix\url{http://link.springer.com/10.1007/s11661-014-2214-4}

\bibitem{Ryo}
R.~Kobayashi, D.~Giofr\'e, T.~Junge, M.~Ceriotti, W.~A. Curtin.
\newblock \href{private communication}{[link]}.
\newline\urlprefix\url{private communication}

\bibitem{Pogatscher2011}
S.~Pogatscher, H.~Antrekowitsch, H.~Leitner, T.~Ebner, P.~Uggowitzer,
  \href{https://doi.org/10.1016%2Fj.actamat.2011.02.010}{Mechanisms controlling
  the artificial aging of al{\textendash}mg{\textendash}si alloys}, Acta
  Materialia 59~(9) (2011) 3352--3363.
\newblock \href {http://dx.doi.org/10.1016/j.actamat.2011.02.010}
  {\path{doi:10.1016/j.actamat.2011.02.010}}.
\newline\urlprefix\url{https://doi.org/10.1016%2Fj.actamat.2011.02.010}

\bibitem{Francis2016}
M.~Francis, W.~Curtin,
  \href{https://doi.org/10.1016%2Fj.actamat.2016.01.014}{Microalloying for the
  controllable delay of precipitate formation in metal alloys}, Acta Materialia
  106 (2016) 117--128.
\newblock \href {http://dx.doi.org/10.1016/j.actamat.2016.01.014}
  {\path{doi:10.1016/j.actamat.2016.01.014}}.
\newline\urlprefix\url{https://doi.org/10.1016%2Fj.actamat.2016.01.014}

\bibitem{zienkiewicz_finite_1977}
O.~C. Zienkiewicz, The finite element method, 3rd Edition, McGraw-Hill, London
  - New York, 1977.

\bibitem{richart2015implementation}
N.~Richart, J.-F. Molinari, Implementation of a parallel finite-element
  library: test case on a non-local continuum damage model, Finite Elements in
  Analysis and Design 100 (2015) 41--46.

\bibitem{ergatoudis1968isoparametric}
J.~G. Ergatoudis, Isoparametric finite elements in two and three dimensional
  stress analysis., Ph.D. thesis, University College of Swansea (1968).

\bibitem{amestoy98mumpsmultifrontal}
P.~Amestoy, I.~Duff, J.-Y. L'Excellent, Mumps multifrontal massively parallel
  solver version 2.0 (1998).

\end{thebibliography}
\end{document}